\documentclass{kluwer}    

\newdisplay{guess}{Conjecture}

\begin{document}                                                                                   
\begin{article}
\begin{opening}         
\title{Three modes of long wavelength flaring from GRS 1915+105: implications for jet formation}
\author{Stephen S. \surname{Eikenberry}}  
\runningauthor{Stephen Eikenberry}
\runningtitle{Modes of Flaring in GRS 1915}
\institute{Cornell University}

\begin{abstract}

	GRS 1915+105 has exhibited at least three modes of
long-wavelength (infrared to radio) flares.  Class A flares are the
bright ($\sim 1$ Jy) radio events whose apparent superluminal motion
in many ways defines the microquasars.  Class B flares have
intermediate ($\sim 50-300$ mJy) flux densities in the IR and radio,
and are associated with hard X-ray dipping behaviors.  Class C flares
are faint ($\sim 5-50$ mJy) IR/radio events associated with soft X-ray
dips.  We discuss each of these classes and their inter-relationships.

\end{abstract}

\end{opening}           

\section{Class A -- Large Radio Flares}

	The discovery of superluminal motions in a radio flare from
GRS 1915+105 by Mirabel \& Rodriguez (1994) in many ways launched the
field of microquasar astrophysics.  Such flares have been seen since
then (e.g. Fender et al., 1999), and always seem to have peak flux
densities of $F_{\nu} \sim 1$ Jy.  The resolved jets with $v = 0.92 c$
show spectral evolution from optically thick to optically thin, and
seem to recur every few days during an ejection episode.  The ejection
episodes themselves recur on timescales of a few years.  While Harmon
et al. (1997) observed an association between the Class A flares and
hard X-ray activity, no simultaneous time-resolved X-ray coverage has
yet been obtained.

\section{Class B -- Mid-size IR/Radio Flares}  

	The second class of long-wavelength events were first clearly
characterized by Fender et al. (1997), whose same-day IR/radio
observations showed flares with flux densities of $F_{\nu} \sim
50-100$ mJy and decay timescales of $\sim 10$ minutes.  The brightness
temperature in the radio implied a synchrotron origin for the flares
and the constancy of the decay timescale over 4 decades of frequency
implied that adiabatic expansion dominates the cooling process.

	Eikenberry et al. (1998a,b) observed a long sequence of Class
B flares in the IR over two nights in August 1997.  These flares had
peak flux densities $F_{\nu} \sim 200-250$ mJy and recurrence
timescales of $\sim 30-60$ minutes.  Simultaneous X-ray observations
with RXTE revealed that these flares are associated with complex X-ray
behaviors (Figure 1).  In particular, the flares come after the hard
X-ray dips, during which the blackbody temperature in the inner disk
drops while the power-law index hardens.  Belloni et al. (1997) first
interpreted dips such as these as the {\it disappearance} of the inner
accretion disk, and combined with the observations of Fender at
al. (1997) and Eikenberry et al., (1998a,b), we recognize that {\bf
these events provide the first direct time-resolved observations of
relativistic jet formation in a black hole system.}

	Mirabel et al. (1998) made simultaneous IR/radio/X-ray
observations of several Class B flares, finding wavelength-dependent
time delays, with the longer wavelengths peaking later.  Such behavior
is roughly consistent with that expected from an expanding van der
Laan blob, and extrapolations based on this timing would indicate that
the ejection occurs near the end of the hard X-ray dip.

\section{Class C -- Faint Flares}

	The existence of a third class of long-wavelength flares was
first clarified through IR/X-ray observations by Eikenberry et
al. (2000).  These flares have peak flux densities in the IR of $\sim
4-8$ mJy and are associated with rapid soft X-ray dips (Figure 2),
during which the blackbody temperature increases and the power-law
index softens (as opposed to the hard Class B dips).  These flares
also seem to explain the quasi-steady IR excess during similar X-ray
behaviors in 1997 (Eikenberry et al., 2000).

	It is interesting to note that the IR flares appear to begin
{\it before} the X-ray activity, implying an ``outside-in''
propagation of the disturbance -- opposite of the Class B behavior.
The spotty coverage from RXTE makes ``aliasing'' a problem for
verifying this.  However, Eikenberry et al. (2000) showed that no
constant delay fits with the X-rays preceeding the IR flares, and even
a non-steady delay would have to reach up to $\sim 3000$ seconds.
Thus, the ``IR-first'' explanation seems the least objectionable for
the observations.

\section{Three Separate Behaviors ?}

	A key issue here is the reality of the distinction between the
three classes presented above.  The Class C flares seem
distinguishable from Class B due to the marked differences in their
corresponding X-ray behaviors (soft versus hard dips, IR-first versus
X-ray-first), as well as differences in recurrence times and
amplitudes.  Class B flares, on the other hand, behave in many ways
like ``baby jet'' versions of Class A ejections.  In particular, the
relative amplitudes and decay timescales are roughly consistent with
identical ejections, but with $\sim 1000$ times less mass in the Class
B events.  However, Dhawan et al. (2000) have used VLBA
to resolve the Class B events, and find that their axis of motion
differs significantly from the Class A events.  Thus, it appears that
there are in fact at least 3 distinct modes of jet production in GRS
1915+105.

\theendnotes


\begin{figure}
\vspace*{85mm}
\includegraphics{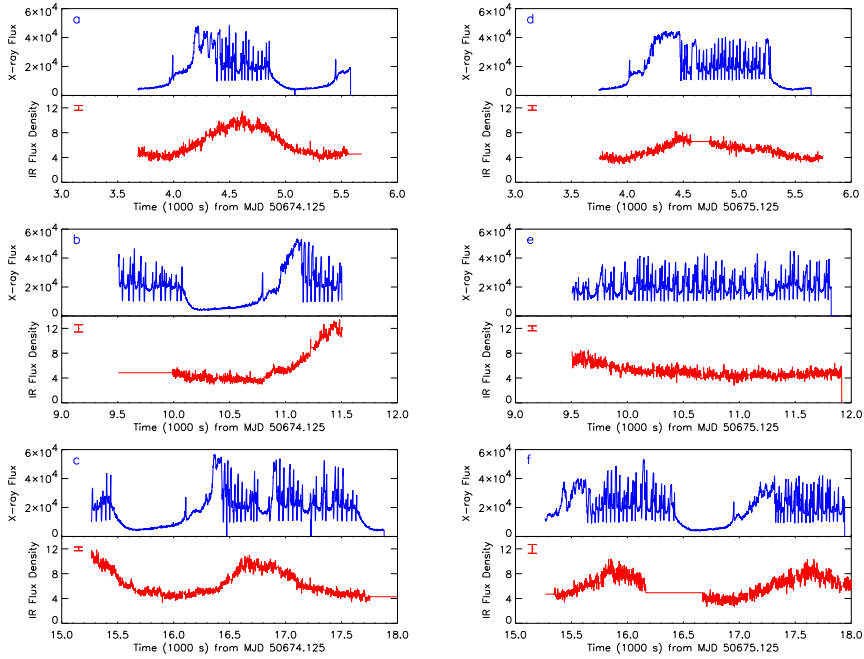}
\caption{Simultaneous IR/X-ray observations of Class B flares from Eikenberry et al. (1998a).}
\end{figure}

\begin{figure}
\vspace*{65mm}
\includegraphics{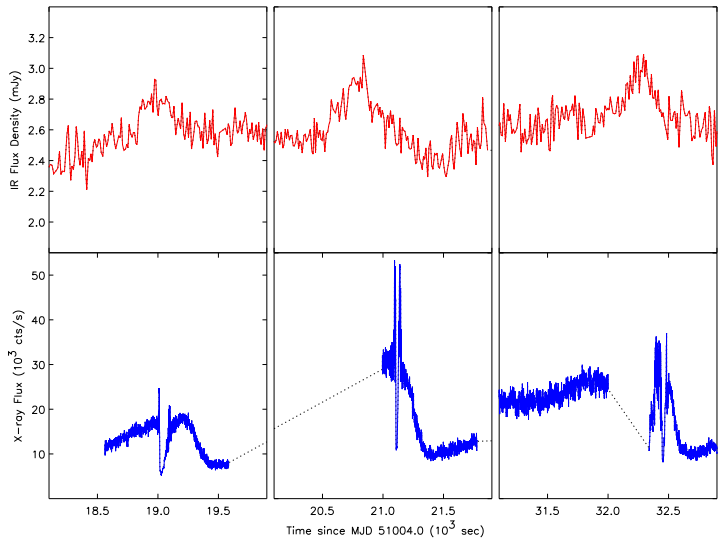}
\caption{Simultaneous IR/X-ray observations of Class C flares from Eikenberry et al. (2000).}
\end{figure}

\end{article}
\end{document}